\documentclass[12pt]{iopart}
\usepackage{booktabs}
\usepackage{graphicx}
\usepackage{amssymb}
\usepackage[utf8]{inputenc}
\usepackage[T1]{fontenc}
\usepackage{color}
\usepackage{hyperref}

\begin{document}

\title{ Teaching gauge theory to first year students.}

\author{N.-E.\ Bomark} 

\address{Institute for natural science, University of Agder, Universitetsveien 25, 4630 Kristiansand, Norway}
\ead{nilserik.bomark@gmail.com}

\begin{abstract}

One of the biggest revelations of 20th century physics, is virtually unheard of outside the inner circles of particle physics. This is the gauge theory, the foundation for how all physical interactions are described and a guiding principle for almost all work on new physics theories. Is it not our duty as physicists
to try and spread this knowledge to a wider audience?

Here, two simple gauge theory models are presented that should be understandable without any advanced mathematics or physics and it is demonstrated how they can be used to show how gauge symmetries are used to construct the standard model of particle physics. This is also used to describe the real reason we need the Higgs field.

Though these concepts are complicated and abstract, it seems possible for at least first year students to understand the main ideas. Since they typically are very interested in cutting edge physics, they do appreciate the effort and enjoy the more detail insight into modern particle physics. These results are certainly encouraging more efforts in this direction.

\end{abstract}


\maketitle

\section{Introduction}

Anyone reading popular science have heard something about quantum physics. There is no shortage of attempts to popularise the intricacies of 
both quantum mechanics and (perhaps to a lesser degree) quantum field theory.

However, when modern physics theories are formulated, perhaps the most crucial ingredient is virtually unheard of outside the particle 
physics community. This is the gauge theory. Without gauge theory the standard model of particle physics cannot be formulated and it is 
fundamentally impossible to fully understand the role of the Higgs field without some understanding of the role of gauge symmetries.

Why is this the case?

This question is hard to answer, gauge theory does not appear any more complicated to understand than quantum physics, however, it may be 
more abstract and it does not have large philosophical implications for how we see the world like quantum physics does. Therefore one could 
argue that teaching gauge theory is a waste since essentially no-one will have any use of it. On the other hand, we are talking about the 
most important (alongside quantum field theory) theoretical development in our understanding of the world since quantum mechanics. Is it acceptable to restrict this knowledge 
to the few who venture into the quite challenging field of advanced quantum field theory?

Just as Shakespeare and Beethoven are considered expected knowledge among the educated populace, it is reasonable to expect knowledgeable 
people and especially physics students and teachers to have some idea about the basis of our current understanding of nature, and for this 
gauge theory is unavoidable.

This may sound all noble, but if this is to actually happen, we must have ways of explaining gauge theory in a not too mathematical manner. 
In the following paper we will explore two models of gauge theories that are designed to introduce the concept without the level of 
abstraction and mathematical complication, present in the particle physics models. Neither of these models were first invented by the 
author, the time-zone model was found at \cite{Gaugeland} and has been elaborated by the author, while the economic model was published by Maldacena~\cite{Maldacena:2014uaa} after \cite{Young} pointed it out, this model has been reformulated to more directly demonstrate the role of symmetries in gauge theory.

The topics presented here are being used to convey an understanding of particle physics to first year students, in a course on particle physics and cosmology. In that course gauge theory is embedded in a wider discussion about what quantum field theory is and it is presented including some more mathematical detail than given here. In this paper the topics are presented as mathematically minimal as possible in order to be accessible to as wide an audience as possible.

In the following we will start with describing what we mean by a symmetry, this is important since the symmetry concept we need is more abstract than most students are used to. In section~\ref{sec:gauge} we introduce the two gauge theory models and explain how gauge theory constitutes the bases for modern physics. The Higgs field is discussed in section~\ref{sec:Higgs}, both why we need it, what it does and how it is connected to gauge theory. Finally in section~\ref{sec:under}, a small enquiry among students about their experiences from being taught this is presented, before we conclude.

\section{Symmetries}
Although the concept of symmetry is familiar to most people, we need to introduce this concept carefully since its mathematical definition is not that obvious and the symmetries we will encounter are rather abstract. The understanding of symmetries may be one of the biggest obstacles in understanding gauge theory.

Mathematically a symmetry is some transformation that leaves the object we are studying unchanged or invariant as well usually express it. In physics the object to remain invariant is usually the laws of physics. Though this definition is technically equivalent with our every day intuition of what a symmetry is, it is formulated in a rather more abstract fashion and thus takes some time getting used to.

To gain experience with this symmetry concept, it is necessary with some examples, unfortunately there are not very many mathematically simple ones. One case is a translational symmetry where a spacial coordinate, $x$, is transformed into, $x\rightarrow x'=x+\Delta x$ where $\Delta x$ is some constant. It can be shown that this is a symmetry to Newton's second law (neither forces nor acceleration are affected by this since $\Delta x$ is a constant and therefore disappear in the derivative). One can also show that Newton's second law respects a rotational symmetry, but that gets more technical since we need a mathematical description of rotations which requires some linear algebra.

One can also point out that electromagnetism is invariant if the electric potential, $V$, is shifted by any constant value, $V\rightarrow V'=V+V_0$, where $V_0$ is any constant number. This should already be known since it is just saying that the level of zero potential can be chosen arbitrarily. This is a nice example since it is part of the gauge symmetry of electromagnetism. For more advanced students one can demonstrate that $V\to V'=V+\frac{d\phi(\vec x,t)}{dt}$ together with $\vec A\to \vec A'=\vec A-\nabla \phi(\vec x,t)$ is a symmetry of Maxwell's equations for all functions $\phi(\vec x,t)$, which is the full gauge symmetry for the electromagnetic field, see \ref{App:QED} for more details.

\section{Gauge theories}\label{sec:gauge}

The key feature to gauge theories, is the concept of a local symmetry. With this we mean that the mathematical transformation that 
defines the symmetry may be applied differently in different points in space. For example, if we have a local rotational symmetry, we rotate 
with a somewhat different angle in different points.

To illustrate how a local symmetry leads to interactions, we shall look at two examples where the symmetry is more intuitive than is the case 
in our particle physics models. Hopefully these models can create an intuition about gauge theory that can be transferred to modern physics 
theories.

\subsection{A time-zone model}
Suppose we try to measure the path of a ball that has been thrown through the air. When doing so it does not matter how we have set the clock 
we use when measuring when the ball is at different positions. We can set the clock forward with 1 hour or back with 15 minutes and nothing 
will change in our results because we only care about time differences.

If you now recall our definition of a symmetry, you recognize that this is exactly that; since we can reset our clock as we like without the 
results being affected, we have a symmetry that we can call a time-zone symmetry.

Let us now assume that we have one clock in each point in space and we use that clock to measure when the ball is at that point. If all 
clocks are synchronized this is no different from the previous situation, we can still reset all clocks as we like as long as we set all of 
them the same way.

To make this a gauge theory we need to make this symmetry local. This means we insist on being allowed to reset all the clocks individually, 
i.e., they will no longer use the same time-zone. Of course, this does not work right away; if all clocks are set randomly with respect to 
each other the time we measure the ball takes between two points has no meaning; how much of this time difference is just due to the clocks 
being set differently?

It does not take too much imagination to solve this, all we need is to keep track of how differently the clocks are set. For this we 
need to introduce a function that we can call $\vec A (x,y,z)$ that tells us the difference between neighbouring clocks. When we reset the 
clocks we update $\vec A (x,y,z)$ with the difference in how much we reset the neighbouring clocks in that point in space, $\vec A (x,y,z)$ is 
a vector because it needs to tell us the difference between the clocks in both the $x$, $y$ and $z$ direction.

We see that by insisting that our system is invariant under this local version of the symmetry we are forced to introduce a vector-field, $\vec A (x,y,z)$. This field we can call a gauge field and we say that we have gauged the symmetry.

If the gauge field, $\vec A (x,y,z)$, is allowed to be dynamical, a fluctuation in this field will act as a force on our ball. To see this imagine that when moving in a certain direction in some region, the gauge field tells us the clocks are set progressively forward; this will make it look like the ball is moving faster since the time differences will be measured as smaller than they actually are. This fluctuation in the gauge field thus gives us an acceleration of the ball, in other words it acts on it with a force.

In summary, by making our time-zone symmetry local, i.e., allowing different time-zones in different points in space, we were forced to introduce a gauge field and that field, when allowed to be dynamical, gives us forces. The point of gauge theory is to use local symmetries to introduce forces, or more generally, interactions into our theories.

\subsection{An economic model}

In principle it is possible to rescale all monetary values with any factor and nothing in the economy would change. If we for example 
multiply all prices with 1 000 000 everything will look very expensive, but if we at the same time multiply all salaries, savings and loans 
with the same number nothing really changed; you can still buy the exact same things for your salary. Of course, in the real world it is not 
that simple, there is psychology involved and people have savings in their mattress, but we will ignore that and pretend these rescalings work 
perfectly.

Again we have a mathematical transformation that leaves everything unchanged, in other words, a symmetry.

The world we live in consists of a number of different countries that all scale their economy differently. How then should we know how to 
deal 
with my Norwegian money when I come to Sweden?

We need a way of keeping track of the differences in scaling between the different countries. This is exactly (at least in our ideal economy) 
what the exchange rates are for. If we also rescale the exchange rates when we rescale everything else in a country we can without 
problem allow every country to maintain whatever scaling they want independently of all other countries.

With the exchange rates in place we see that our scaling symmetry is now local, we can scale the economy of every country independently of 
the other countries and the exchange rates is our gauge field that allow this local symmetry.

Let us now imagine that the exchange rates fluctuate and perhaps we notice that if I start with 100 Norwegian kroner, if I go to Sweden I get 
110 Swedish kroner for those. If I then take my 110 Swedish kroner and go to Denmark I get only 88 Danish kroner, but for those I get 104.5 
Norwegian kroner. This means that I have more money than I started with and if this is the case, investors will travel this circle over and 
over again to make money. Like with the time-zone model the fluctuation in the gauge field produces a force that makes investors move 
around in this circle.

In this example we can also see that the interaction goes both ways, we know from the real world that movements of money between countries will change the exchange rates. In other words, the gauge field (exchange rates) affect the money and the money affect the gauge field. This bidirectionality of interactions is 
a general feature of physics theories; if something affects something else, the second thing also affects the first.

\subsection{The importance of gauge theory in modern physics}

When we summarise the above two models of gauge theories, we see that the main point is to take a global symmetry and make it local, i.e., 
allow different transformations in different points in space, and then, to ensure the theory to respect this local symmetry, we need to 
introduce gauge fields that related the symmetry transformations in neighbouring points. Allowing the gauge fields to be dynamic means we get 
interactions between the gauge fields and the other components in our theory.

To use this on physics theories we need to figure out which symmetries to start with, in particle physics these symmetries will usually be 
rather abstract. The simplest example is Quantum ElectroDynamics (QED) where the symmetry transformation consist of multiplying the 
fields\footnote{Despite its name, particle physics does not foremost deal with particles, but with fields. We have for example a field that 
represents all electrons and positrons in the universe. This means all fields have particles associated with them and all particles have corresponding fields.} 
corresponding to charged particles with an arbitrary complex phase\footnote{This means multiplying the fields with a factor $e^{i\theta}$, which is the same as rotating in the complex plane with an angle $\theta$. This is described in some mathematical detail in \ref{App:QED}.}. This symmetry is harder to visualise but 
is similar to the time zone symmetry in that the transformation is parametrised by an angle; reading a clock can also be viewed as measuring 
an angle.

In making this symmetry local we get a gauge field, usually written $A_\mu$, that turns out to be the electromagnetic field and we have 
indeed deduced electromagnetism. This is actually rather remarkable, it took physicists almost 100 years to get the theory of 
electromagnetism 
right and here we can deduce it in a few lines of algebra from just making this phase symmetry local!

Today all interactions are described by gauge theories. As a matter of fact, this is the only way we found to describe the forces we see in 
nature in a mathematically consistent way\footnote{This may sound surprising, why should mathematical consistency be hard to achieve? \\ 
Because when one field interacts with another, the properties of the first changes as a result of the interaction. In some cases these 
changes lead to probabilities being larger than one and hence the theory being mathematically inconsistent. Gauge symmetries prevent 
this from happening.}. Let us see which symmetries are used for the various interactions.

\paragraph{Electromagnetism ---}as already mentioned a phase rotation for all electrically charged fields. This gives us the mathematically 
simplest gauge theory. The gauge field is the electromagnetic field whose quanta are called photons. For reference, this theory is presented in some mathematical detail in \ref{App:QED}.

\paragraph{Weak nuclear force ---}here we have a more complicated symmetry that we can see as two independent phase rotations, one 
for electrons and down quarks and one for neutrinos and up quarks, and one transformation that transforms electrons into neutrinos and down 
quarks into up quarks and vice verse. Since there are three components to this symmetry, we get three gauge fields, the $W^+$, $W^-$ and $Z$.

\paragraph{Strong nuclear force ---}again a more complicated symmetry, this time it transforms the quarks of different colour charge into 
each other and rotates their phases. Though not obvious, it turns out that this contains eight independent transformations and hence we get 
eight gauge fields called gluons.

\paragraph{Gravity ---}though not part of particle physics since we have not managed to write down a mathematically consistent quantum 
field theory for it, gravity is actually also a gauge theory. The symmetries here are more intuitive since they are symmetries in our physical 
space-time; these include rotational invariance (the laws of physics are the same in all directions), translational invariance (the laws of 
physics are the same everywhere) and Lorentz invariance (the laws of physics are the same in all inertial coordinate-systems). The gauge 
fields here are basically the curved spacetime itself and if we manage to quantize this its quanta would be gravitons.

We see that gauge theories are literally the foundation of our whole understanding of physics; in the vocabulary of contemporary physicists 
an interaction or theory is defined by which gauge symmetries it is based on. This is why it is such a shame that knowledge about gauge 
theory is limited to those few with Master degrees in particle physics.

\section{The Higgs mechanism}\label{sec:Higgs}

One of the most important events of the 21st century took place at CERN in 2012 when the Higgs boson was discovered. However, to understand 
why this was so important is not easy. Most popular accounts say something about explaining why particles have mass or that Higgs gives mass 
to other particles.

Though it is true that the other particles of the standard model get their masses from interacting with the Higgs, this does not really get 
to the core of the issue. Would it not be easier to just do what we always have done and treat mass as an intrinsic property that just happen 
to have the values we measure?

One could object it would be more satisfying to understand where those values come from. However, the Higgs mechanism does not help in that, 
it does not explain any deeper meaning of the concept of mass and it does not tell us what the values of the various particle masses should 
be, they are still just numbers we take from measurements. As a matter of fact, if we could keep using the measured mass-values without 
invoking the Higgs, we would.

What is then the purpose of the Higgs field?

It is all about the weak nuclear force. We can immediately see that there is something odd with it; we learned that it is based on a gauge 
symmetry that amongst other things transforms electrons into neutrinos and vice verse. But how can that possibly be a symmetry? If all 
electrons in our atoms were replaced by neutrinos, things would certainly change a lot. Something is wrong here. Contrast this with the strong nuclear force that transform for instance red upquarks to blue upquarks; there is a reason they are both called upquarks, they are identical except for the colorcharge so swapping them makes no difference whatsoever.

There are other issues with the weak force as well, the gauge fields $W^+$, $W^-$ and $Z$ are known to have large masses; this is why they 
have so little impact on our lives, they cause $\beta$-decay and that is essentially it. However, massive gauge fields are problematic, it turns out they break the 
gauge symmetry (this is not obvious, some details are given in the end of \ref{App:QED}).

We are in trouble, the bases for the theory is the gauge symmetry and that symmetry must be maintained for it to be mathematically 
consistent, but our observations tell us that the symmetry is broken. What are we to do?

Here comes Higgs to the rescue!

\subsection{Spontaneous symmetry breaking}

The trick is to make it look like the symmetry is broken without actually breaking it.

This is not as strange as it first may sound, we have plenty of everyday examples of this happening. The most obvious is the direction down; 
we know that the laws of physics are the same in all directions, but since we happen to live on a big planet it seems that the direction down 
is special. We can say that the symmetry between all directions is spontaneously broken by the gravitational field of the Earth. The laws of 
physics still respect this symmetry, but here on the Earth’s surface we do not really see it.

A more direct analogy is a ferromagnetic material where at high temperature all the little magnets (spin of the electrons) are pointing in 
random directions and hence all directions look the same. At low temperature the magnets arrange themselves so that they all point in the 
same direction. This direction now looks special and the directional symmetry seems to be broken, but it is only the lowest energy state that 
breaks the symmetry, it is still present in the fundamental equations governing the system.

\begin{figure}%
\includegraphics[width=\columnwidth]{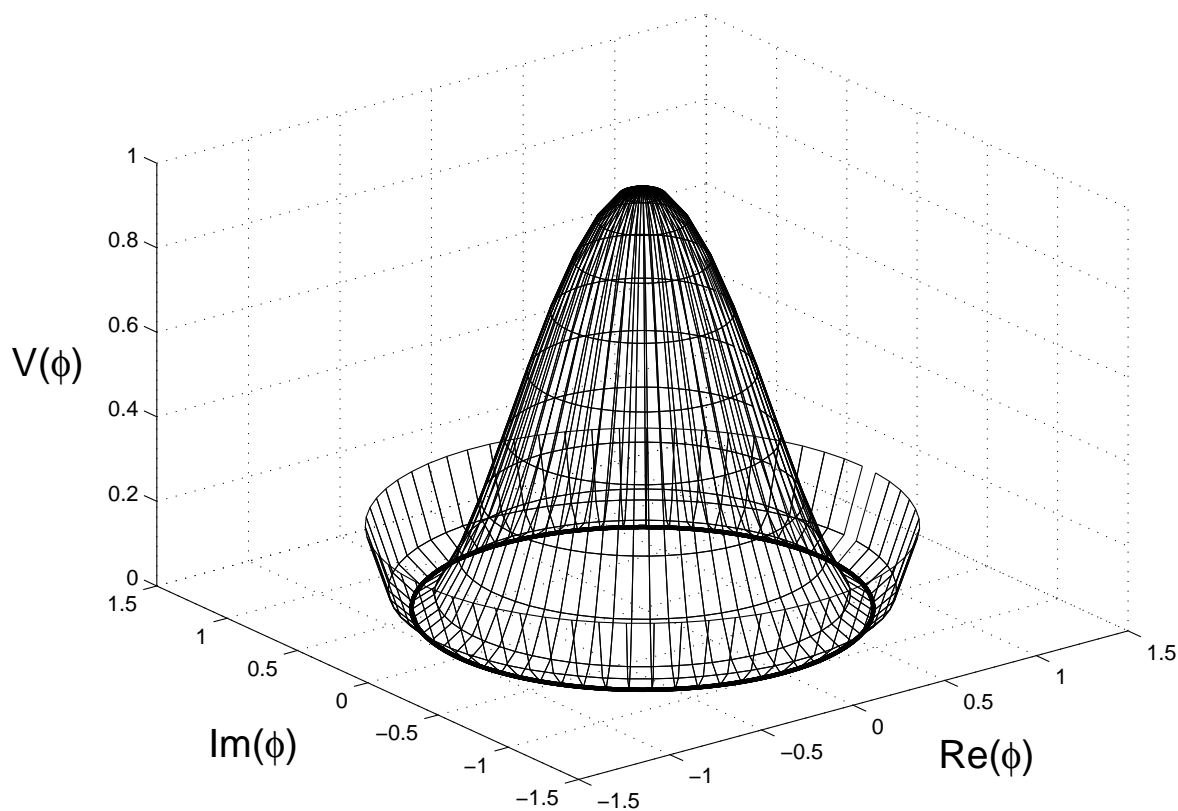}%
\caption{A rotationally symmetric potential, with a minimum that spontaneously breaks the symmetry. In the vacuum state at the bottom of the potential, we do not see the rotational symmetry, things do not look the same in all directions from down there.}%
\label{fig:potential}%
\end{figure}

We need to do the same with the gauge symmetry of the weak nuclear force. To do this we need a scalar field (the Higgs field) and then we 
give it a potential energy such that the lowest energy does not happen in a symmetric point. Such a potential is depicted in figure~\ref{fig:potential}.

It is not easy to comprehend what it means for a symmetry to be spontaneously broken; in exactly what sense is it broken? 

We can understand this intuitively by looking at figure~\ref{fig:potential}, the potential is rotationally symmetric which here means we have a rotational symmetry. It should be noted that this rotational symmetry does not rotate real space, but field space, the space where the Higgs field lives, notice that the axes display the real and imaginary part of the Higgs field. This symmetry is in fact the type of symmetry responsible for electromagnetism, i.e., a phase rotation symmetry. Why not use the actual symmetry of the weak force? Because then the figure would need to be five dimensional.

A state of lowest energy (the vacuum state) will sit at the bottom of the potential and from there it does not look rotationally symmetric any more (the bottom is in a valley with two flat directions and two walls). In more technical terms we can say that if we rotated everything around the vacuum state, the theory is no longer invariant. If we on the other hand were to rotate everything including the vacuum state around the origin, the theory is invariant and therefore the symmetry is still there, it just does not look like that from our vacuum state. This is what it means to be spontaneously broken.

\subsection{Higgs and mass}

In short, the Higgs field is included in the standard model to spontaneously break the gauge symmetry of the weak nuclear force. In doing so it turns out the Higgs gives mass to the $W^+$, $W^-$ and $Z$ bosons, the masses that would otherwise be forbidden by the gauge symmetry. This is the main reason the Higgs is needed.

The masses of other particles are more complicated, it turns out that also all the fermions that interact with the weak nuclear force (i.e., all known fermions) need the Higgs field to get their masses. This fact is far from trivial\footnote{Fermion masses mix left and right chiral states, but the weak force only sees left chiral states and therefore forbids those masses. It is hard to formulate this in a simple way.}, but it means all the masses present in the standard model comes from the Higgs field\footnote{As is often pointed out; the mass of the proton does not, it mostly comes from the strong nuclear force, but we talk about fundamental particles here.}. As a consequence, how massive a particle is is determined by how much it interacts with the Higgs field and all masses are proportional to the absolute value of the Higgs field in the vacuum state, i.e., the radius of the circle at the bottom of the potential in figure~\ref{fig:potential}.

We now see why everyone keep talking about the Higgs giving mass to particles, but it is not about explaining what mass is or why particles have mass or the values of those masses. It is rather that the Higgs mechanism allows the masses to be included without the mathematics breaking down. If you ask where that strange potential energy in figure~\ref{fig:potential} comes from, we have no answer, or if you ask why the masses are what they are, we can say it is determined by their interactions with the Higgs field, but what determines that interaction? We do not know, so the question is not answered, merely reformulated.

The standard model cannot be formulate consistently without the Higgs field, that is why it is so important, but it does not answer any deeper questions about mass.

One may also wonder whether for instance Dark Matter needs the Higgs field to get its mass; here the answer is most likely no. We have seen that the particles whose mass is forbidden by the gauge symmetry of the weak nuclear force need the Higgs field for their mass, but if there is a particle that for instance do not interact with the weak force or does so in a way that does not forbid a mass, that particle can have mass without any interaction with the Higgs field. 
It just so happens that all known particle masses are forbidden by the weak force gauge symmetry, with one exception; the Higgs field itself. There may well be more exceptions we do not know about and most candidates for Dark Matter fall in this category.

\section{Some observations from teaching gauge theory}\label{sec:under}

These models have been used to teach gauge theory to first year physics students in a course in particle physics and cosmology at the University of Agder in Kristiansand, Norway. It was first used in 2016 and the curriculum has been developed since then. To gauge the reception of this, the students of 2020 answered some questions about their experience with these concepts and here we present the results of the enquiry.

The students answered a short electronic questionnaire about their thoughts about learning about gauge theory and spontaneous symmetry 
breaking. These questions were not designed to test their understanding of gauge theory or if they really see the underlying structure of 
the standard model, that would require a much more elaborate investigation that the circumstances did not allow.

The question that this investigation tries to answer is: \\ {\it What are the students thoughts on whether teaching gauge theory on this 
level is worthwhile?}

\subsection{The students}

All the students were at the end of a one year physics program. This one year is all the physics they have studied at university and include mechanics, 
thermodynamics, electromagnetism, quantum physics, wave mechanics and the particle physics and cosmology course under discussion here. Some 
of the students are to become high school teachers in mathematics and physics, some take this as part in their bachelor in mathematics; both 
these groups have had one year of mathematics at the university prior to the year of physics. There are also students who start the physics 
year without any university level education in either mathematics or physics, among those are one philosophy student and several students coming directly from high school.

It should be emphasised that the particle physics and cosmology course does contain a lot more than just gauge theory, it also covers special relativity, nuclear physics, gravity and a qualitative description of general relativity, in addition to an overview of particle physics and cosmology. The course also includes some discussions about what quantum field theory is (in non-technical terms) and the discussion about gauge theory is a bit more elaborate than presented here, but the essence is the same. The students are given some more mathematical details, but are not expected to reproduce that on an exam.

There were eleven students that took the exam in the course and six of those answered the questionnaire. Three of the students are female and the rest 
male, and the age ranges from late teens to middle age. All the students are fluent in Norwegian (the questionnaire was in Norwegian), but two 
have other mother tongues. It is not known who among students answered the questions, so it is possible that the distribution among those 
students differ from the numbers given above.

The questionnaire was completely voluntary and anonymous and this study has been reported to the Norwegian Centre for Researchdata (NSD). It should also be pointed out that this course in 2020 was taught almost entirely digitally. With such a small class it might not have made a very big difference, but it limits the possibility for discussions with the students somewhat.

\subsection{Questions}
To ensure as high degree of participation as possible the questionnaire was kept very short with just five questions of which four were multiple choice. All the questions as well as the rest of the course were given in Norwegian, so in the following, replies will be translated to English but the original Norwegian text will also be given for reference.

All the questions, both in original and translated are given in \ref{App:questions}. Since the number of questions and student replies are rather limited, the data analysis is restricted to showing the distribution of answers on the multiple choice ones and quote the text replies to the last question.

This questionnaire was given after the exam in the course and the students were explicitly encouraged to answer what they think rather than what they think the author wanted to hear.

Though most questions were designed to only find out what the students think about the topic rather than how much they understand of it, question 3 is an attempt to check that they gathered what the purpose of the Higgs mechanism is, this is done so that the replies about the purposefulness of teaching about it can be trusted; if the students have misunderstood what the purpose of the Higgs mechanism is, one cannot take comments about whether it is worth teaching too seriously. It is, though, a valid question as to what extent question 3 succeeds in this.

\subsection{Results and discussion}

The replies to the multiple choice questions are summarised in table~\ref{tab:Svar}. We see that the answers are very well aligned among the students and only a few have differing opinions. The replies are also encouragingly positive, though perhaps one should bear in mind that not everyone answered and it is possible that those who did not answer are less positive. It is also possible that it feels easier to give a positive reply if one does not have very strong feelings against that. Therefore one should perhaps be a little careful in interpreting the result, however we do not have any concrete reasons why the non-responding students should be less positive than the responding nor the responding ones less positive than they express, so this is indeed a positive result.

\begin{table}
	\centering
		\begin{tabular}{|p{12cm}|l|}
		  \hline
		  Option & Replies \\\hline
			Question 1. & \\\hline
			It makes it possible to see the whole picture of the standard model. & 5 \\
			It is interesting to get some insight into how modern physics theories are constructed. & 4 \\
			It gets mostly confusing since we do not have the background knowledge required to understand it properly. & 1 \\\hline
			Question 2. & \\\hline
			Both models are good and it is good to have several examples. & 4 \\
			I liked the economic model best. & 1 \\\hline
			Question 3. & \\\hline
			Yes, we cannot formulate the standard model without spontaneous symmetry breaking. & 5 \\
			It is not strange that people ponder about what mass is and why the electron is lighter than the proton. & 1 \\\hline
			Question 4. & \\\hline
			Yes, I now have a much clearer picture of what it is about. & 3 \\
			Yes, I have gotten insight into a physics-world I did not know existed. & 2 \\
			No, but the details are somewhat clearer. & 1 \\\hline
		\end{tabular}
	\caption{Summary of the replies to the multiple choice questions. Note that on question 1. more than one reply was allowed, hence the larger count.}
	\label{tab:Svar}
\end{table}

Let us look at each question a bit closer, starting from the beginning with question 1. The purpose of this question was to see if the students think learning gauge theory makes the standard model more of a unity rather than just a collection of particles. From the replies they seem to agree that it does, hence this argument for teaching the subject might have some merit, though one student thinks it is hard without more background knowledge; a valid point that needs to be taken seriously, but since only one student answered this it does not seem to be a big problem. We also see that many of the students appreciate getting some insight into how modern physics theories are constructed, this is also very gratifying to hear as this is part of the reason for this whole endeavour.

Question 2 was designed to check how the models worked and it seems the students liked having two somewhat different models. The positive response here suggests that this way of describing gauge theory has some merit.

The third question was a check to see if the students have a better understanding of why the Higgs is so important than the ``we need to understand mass'' that dominates the popular science. It seems five out of six students have, though of course, as always with these sort of questions it is impossible to know to what degree they understand this or if they just memorised a sentence. We see that one student went for the ``explain mass and why electron is lighter than proton'' option and as should be clear by now, this is just not why the Higgs is included in the standard model. If this would have been the prevalent answer it would have been worrying, but with only one student we should not pay too much attention to it, though it might indicate room for improvement in the teaching.

The last multiple choice question concerns the students ideas of what modern particle physics is all about. The overall response indicate that the goals with the teaching has been to some degree accomplished, the students think they now better see what it is about and two say they have gotten insight into a physics world they did not know existed, since this was part of the reason to teach gauge theory in the first place, this is a very positive result.

Let us finally look at the replies to the open question in the end. Since only five students wrote something, all the replies are listed below with some brief comments. The question was whether they think it makes sense to learn gauge theory and the Higgs mechanism at this level.

\begin{quote}
Ja, synest det var spesielt spennende med Higgs. Jeg hadde hørt mye om ``Guds partikkelen'' da den ble oppdaget, men hadde ikke gått inn i 
dybden. {\it Yes, I think it was especially exciting with Higgs. I had heard a lot about ``the God particle'' when it was discovered, but had not studied it in depth.}
\end{quote}

\noindent This illustrates a point, many physics students are very interested in physics and read popular science about it. It therefore makes sense that they also find it interesting learn more about these things, even if it is not directly useful from a practical point of view. The God particle is a nickname for the Higgs which is popular in the media, but never used by particle physicists.

\begin{quote}
 absolutt riktig {\it absolutely the right thing to do}
\end{quote}

\noindent Clearly positive response.

\begin{quote}
Ja, det er komplisert og det kreves å sette seg ned å jobbe i litt ekstra, men det lønner seg i det lange løp. {\it Yes, it is complicated and require some extra work, but it pays of in the long run.}
\end{quote}

\noindent It is nice to see that something being difficult is not necessarily seen as negative by students. Working hard to learn complicated things have its benefits by itself.

\begin{quote}
Jeg har personlig vært interessert i fysikk i mange år, men aldri hatt noe bedre forståelse enn den lille man klarer å opparbeide seg gjennom 
populærvitenskapen. Å få lære om Gauge-teori og Higgs-feltet, selv om vi har fysikk på et relativt lavt nivå, har gitt meg et annet syn på 
hva partikkelfysikk er. Jeg synes det er viktig å få lære litt om disse tingene for å få et riktigere bilde av partikkelfysikk og av 
universets utvikling. Jeg synes også det er god moral sånn vitenskaplig å tenke at feltet ikke er helt låst for de aller flinkeste, men at 
alle med interesse for faget kan dra noe nytte ut av forenklede forklaringer. Selv om jeg ikke skal studere mer fysikk kommer jeg til å 
fortsette å prøve å forstå partikkelfysikken bedre, og jeg tror også jeg ville synes det var kjedelig om jeg ikke fikk en oversikt over hva 
man kan dykke ned i. {\it I have personally been interested in physics for many years, but never had any better understanding than the little you can get from popular science. To learn about gauge theory and the Higgs field, even though we have physics at a rather low level, have given me a different view on what particle physics is. I think it is important to learn something about these things to get a more correct picture of particle physics and the history of the universe. I also think it is good moral, scientifically, to think that the field is not locked for the very best, but that everyone with an interest for the subject can benefit from simplified explanations. Even though I will not study more physics, I will continue to try to understand particle physics better, and I also think I would find it sad if I did not get an overview of what one can dive into.}
\end{quote}

\noindent This is as if I would have written it myself; it is very satisfying that a student practically gives the exact arguments that motivated this effort in the first place.

\begin{quote}
Ja det gir mening, men kan være litt forvirrende å tenke seg hvilke oppgaver som kan være aktuelle for temaet. {\it Yes it makes sense, but can be a bit confusing to imagine what questions can be relevant for the subject.}
\end{quote}

\noindent This statement mostly concerns the exam in the course, and it is true that it is hard to make good exam question on gauge theory at this level, but that is a different question as to whether we should include it in the curriculum.

\subsection{Summary of the student responses}
All in all, the students agree that gauge theory and the Higgs mechanism are interesting topics that they want to learn about. It is hard to learn, but that should not discourage us from trying.

The responses contain many of the arguments used to motivate this effort; the desire to understand things heard about in popular science in more detail, to get a better understanding about what particle physics is about and to see the underlying structure of the standard model. This can be taken as an indication that these efforts are on the right track.

\section{Conclusions}

In this work it has been demonstrated how the concept of gauge theory and reasons behind the Higgs mechanism can be described with relatively little mathematical complexity. This opens the door for a broader audience to get a more firm understanding of the theoretical underpinnings of modern physics.

Though these concepts are far from simple, it is fully possible to teach to first year bachelor students and possibly also earlier than that if required. This is of course also of interest to the scientifically interested populace outside the universities.

A small questionnaire among the students exposed to this teaching, indicates a significant interest in this and in general a positive reception. We can take this as a confirmation that this sort of effort is appreciated and these simple models can hopefully serve as part of an increased effort by the particle physics community to give the interested audience a more detailed and correct description of what we are doing.


\appendix

\section{The questionnaire}\label{App:questions}

Here follow the questions asked to the students. Since they were given in Norwegian, the original Norwegian texts is given together with an English translation in italics. The first four questions are multiple choice where the first allow several options while the other three only allow one option to be marked. The last question is an open one where the students were allowed to write down their own thoughts.

\renewcommand{\labelitemi}{$\square$}	

\noindent { \bf 1. Hva syns du om å lære om gauge teori? Flere svar mulige. {\it What do you think about learning gauge theory? Multiple answers possible.}}

\begin{itemize}
	\item Det gjør det mulig å se helheten i standardmodellen. {\it It makes it possible to see the whole picture of the standard model.}
	\item Det er interessant å få en innblikk i hvordan moderne fysikk-teorier er bygget opp. {\it It is interesting to get some insight into how modern physics theories are constructed.}
	\item Det blir mest forvirrende siden vi ikke har nok bakgrunnskunnskap til å forstå det ordentlig. {\it It gets mostly confusing since we do not have the background knowledge required to understand it properly.}
	\item Det finnes mer fornuftige ting å bruke tiden på. {\it There are better ways to use the time.}
	\item Det er vanskelig å se poenget med det. {\it It is hard to see the point.}
\end{itemize}

\renewcommand{\labelitemi}{$\circ$}

\noindent { \bf 2. Hva syns du om våre modeller for gauge-teorier; eksemplet med forskjellige tidssoner og den økonomiske modellen?	{\it What do you think about our models for gauge theories; the example with different time-zones and the economic model?}}

\begin{itemize}
	\item Jeg likte tidssonen best. {\it I liked the time-zone model best.}
	\item Jeg likte den økonomiske modellen best. {\it I liked the economic model best.}
	\item Begge modellene er gode og det er greit med flere eksempler. {\it Both models are good and it is good to have several examples.}
	\item De kompletterer hverandre godt. {\it They complement each other well.}
	\item Jeg skjønte aldri helt hva vi skal lære fra noen av dem. {\it I never understood what we were supposed to learn from any of them.}
	\item Det går greit å forstå modellene, men jeg skjønner ikke hvordan det er relatert til partikkelfysikk. {\it It is okay to understand the models, but I do not understand how they are related to particle physics.}
	\item Jeg syns det går like greit å forstå gauge-teori uten de modellene. {\it I think it is just as easy to understand gauge theory without those models.}
\end{itemize}

\noindent { \bf 3. Syns du at du fått en forståelse for hvorfor Higgs er så viktig for standardmodellen? {\it Do you think you have understood why the Higgs is so important for the standard model?}}

\begin{itemize}
	\item Ja, vi kan ikke formulere standardmodellen uten spontant symmetribrudd. {\it Yes, we cannot formulate the standard model without spontaneous symmetry breaking.}
	\item Det er jo ikke rart at folk spekulert i hva masse er og hvorfor elektroner er så mye lettere enn protoner. {\it It is not strange that people ponder about what mass is and why the electron is lighter than the proton.}
	\item Jeg skjønner ikke helt hvorfor man ikke bare kan si at partiklene har de masser de har. {\it I do not really understand why we cannot just say that the particles have the masses they have.}
	\item Jeg skjønte aldri helt hvorfor vi trengte Higgs. {\it I never really understood why we need the Higgs.}
	\item Konseptet spontant symmetribrudd var vanskelig å forstå. { \it The concept of spontaneous symmetry breaking was hard to understand.}
	\item Nei, denne delen skjønte jeg ikke helt. {\it No, this part I did not really understood.}
\end{itemize}

\noindent { \bf 4. Har din forståelse av hva moderne partikkelfysikk dreier seg om endret seg som følge av undervisningen i gauge-teori og Higgs-mekanismen?{\it Have your understanding of what modern particle physics is about, changed as a consequence of the teaching of gauge theory and the Higgs mechanism?}}

\begin{itemize}
	\item Ja, jeg har nå en mye klarere bild av hva det dreier seg om. {\it Yes, I now have a much clearer picture of what it is about.}
	\item Ja, jeg har fått innblikk i en fysikk-verden som jeg ikke visste eksisterte. {\it Yes, I have gotten insight into a physics-world I did not know existed.}
	\item Jeg hadde egentlig ikke noen ide om hva partikkelfysikk er, nå skjønner jeg noe av hva det dreier seg om. {\it I did not really have any idea what particle physics is, now I have some understanding of what it is about.}
	\item Jeg visste veldig lite om partikkelfysikk, og er fortsatt ganske usikker på hva det egentlig dreier seg om. {\it I knew very little about particle physics, and I am still uncertain about what it is.}
	\item Nei, Ikke noe særlig. {\it No, not particularly.}
	\item Nei, men detaljene er noe klarere. {\it No, but the details are somewhat clearer.}
	\item Nei, jeg føler jeg visste ganske godt hva partikkelfysikk er allerede før kurset. {\it No, I feel I knew quite well before the course what particle physics is.}
	\item Ja, litt. {\it Yes, a little}
\end{itemize}

\noindent { \bf 5. Syns du det gir mening å lære om disse tingene (gauge-teori og Higgs) på dette nivået? {\it Do you think it makes sense to learn about these things (gauge theory and the Higgs) at this level?}}

\section{QED}\label{App:QED}

For reference and more advanced students, let us deduce the Lagrangian for Quantum ElectroDynamics (QED). In the following the reader is assumed to be familiar with four-vector notation and Lagrangians.

We start with a free fermion field $\psi$. The free field Lagrangian is,

\begin{equation}\label{eq:FriLag}
 \mathcal L = \bar \psi (i\gamma^\mu\partial_\mu-m)\psi,
\end{equation}
where $\gamma^\mu$ are the gamma matrices, for our purposes they can be considered some constants, $\partial_\mu$ represents derivatives in space and time and $m$ is the mass of the fermions.

This Lagrangian has a global symmetry $\psi\to \psi'=e^{i\phi}\psi$. Since $\bar \psi$ is a kind of complex conjugate, it must transform like, $\bar \psi\to \bar \psi'=e^{-i\phi}\bar \psi$, and then the two exponentials cancel in the Lagrangian, which therefore remains invariant.

Let us try a local version, $\psi\to \psi'=e^{i\phi(\vec x,t)}\psi$. This no longer leaves the Lagrangian unchanged since we cannot take $e^{i\phi(\vec x,t)}$ through the derivative $\partial_\mu$, we must use the chain rule. Doing so yields,

\begin{eqnarray}
	\nonumber \mathcal L\to \mathcal L' &= \bar \psi e^{-i\phi(\vec x,t)} (i\gamma^\mu\partial_\mu-m)e^{i\phi(\vec x,t)}\psi =\\
	&= \bar \psi  (-\gamma^\mu \partial_\mu \phi(\vec x,t)+i\gamma^\mu\partial_\mu-m)\psi\neq  \mathcal L.
\end{eqnarray}

To restore the invariance we must change the partial derivative to a covariant derivative $D_\mu \equiv \partial_\mu+ieA_\mu$, where $e$ is the elementary charge and $A_\mu$ is a vector-field that transforms according to $A_\mu\to A'_\mu= A_\mu-\frac{1}{e}\partial_\mu \phi(\vec x,t)$, under these symmetry transformations. One can then show that,
\begin{equation}\label{eq:LagQED}
 \mathcal L = \bar \psi (i\gamma^\mu\partial_\mu-\gamma^\mu eA_\mu-m)\psi-\frac{1}{4}F_{\mu\nu}F^{\mu\nu},
\end{equation}
where $F_{\mu\nu}\equiv \partial_\mu A_\nu-\partial_\nu A_\mu$ is a kind of kinetic energy to the field $A_\mu$, is invariant under this gauge symmetry. 

It turns out that (\ref{eq:LagQED}) describes all we know about electromagnetism. We have deduced one of the most important theories in human history with nothing more than insisting that the phase symmetry, $\psi\to \psi'=e^{i\phi}\psi$, should become a local symmetry.

To see why a mass term for the gauge field is problematic, let us look at such a term (we do not want a mass term for the photon, but we use this as a simple example to illustrate the point). Such a term has the form $\frac{1}{2} m^2A^\mu A_\mu$, and we see that this cannot in general be invariant under the symmetry transformation $A_\mu\to A'_\mu= A_\mu-\frac{1}{e}\partial_\mu \phi(\vec x,t)$, i.e., $\frac{1}{2} m^2A^\mu A_\mu \neq \frac{1}{2} m^2A'^\mu A'_\mu$. That is why gauge fields cannot have mass without the gauge symmetry being spontaneously broken.

\section*{References}

\end{document}